\documentclass[prd, twocolumn, longbibliography, superscriptaddress, nofootinbib, floatfix]{revtex4-2}

\pdfoutput=1
\usepackage{amsmath,amssymb,amsfonts}
\usepackage{mathrsfs}
\usepackage{bbm}
\usepackage{slashed}
\usepackage{graphicx}
\usepackage{verbatim}
\usepackage[T1]{fontenc}
\usepackage[utf8]{inputenc}
\usepackage[colorlinks]{hyperref}
\usepackage{mathbbol}
\usepackage[dvipsnames]{xcolor}
\usepackage[normalem]{ulem}
\usepackage{svg}
\usepackage{lipsum, babel}
\usepackage{physics}
\usepackage{graphicx}% Include figure files
\usepackage{dcolumn}% Align table columns on decimal point
\usepackage{bm}% bold math
\usepackage{tensor}
\usepackage{comment}
\usepackage[makeroom]{cancel}

\newcommand{\be}{\begin{equation}}
\newcommand{\ee}{\end{equation}}

\newcommand{\cC}{\mathcal{C}}
\newcommand{\cQ}{\mathcal{Q}}
\newcommand{\cR}{\mathcal{R}}

\newcommand{\Db}{{\bar{D}}}
\newcommand{\Rb}{{\bar{R}}}
\newcommand{\gb}{{\bar{g}}}

\newcommand{\xt}{{\tilde{x}}}

\DeclareMathOperator{\STr}{STr}

\begin{document}

\title{On the reconstruction problem in Quantum Gravity}

\author{Mathijs Fraaije}
\email[]{mfraaije@science.ru.nl}
\affiliation{
Institute for Mathematics, Astrophysics and Particle Physics (IMAPP),Radboud University Nijmegen, Heyendaalseweg 135, 6525 AJ Nijmegen, The Netherlands
}
\author{Alessia Platania}
\email[]{aplatania@perimeterinstitute.ca}
\affiliation{
Perimeter Institute for Theoretical Physics, 31 Caroline Street North, Waterloo, ON N2L 2Y5, Canada
}
\author{Frank Saueressig}
\email[]{f.saueressig@science.ru.nl}
\affiliation{
Institute for Mathematics, Astrophysics and Particle Physics (IMAPP),Radboud University Nijmegen, Heyendaalseweg 135, 6525 AJ Nijmegen, The Netherlands
}

\begin{abstract}
Path integrals and the Wilsonian renormalization group provide two complementary computational tools for investigating continuum approaches to quantum gravity. The starting points of these constructions utilize a bare action and a fixed point of the renormalization group flow, respectively. While it is clear that there should be a connection between these ingredients, their relation is far from trivial. This results in the so-called reconstruction problem. In this work, we demonstrate that the map between these two formulations does not generate non-localities at quadratic order in the background curvature. At this level, the bare action in the path integral and the fixed-point action obtained from the Wilsonian renormalization group differ by local terms only. This conclusion does not apply to theories coming with a physical ultraviolet cutoff or a fundamental non-locality scale.
\end{abstract}


\maketitle

%\tableofcontents
%------------------------------------------------------
\section{Introduction}
%------------------------------------------------------
Finding a consistent and predictive quantum theory for the gravitational interactions is a longstanding problem in theoretical physics. One way to approach this problem is to make sense of the path integral over 
spacetime metrics \cite{Kiefer:2014sfr,Mottola:1995sj,Loll:2022ibq}. In Euclidean signature this reads 
\be\label{eq:pi}
Z \equiv \int \mathcal{D}g \,  e^{- S^{\rm bare}[g]} \, .
\ee
Here $\mathcal{D}g$ is a measure on the space of geometries and the functional $S^{\rm bare}[g]$ is the bare action providing the statistical weight for each configuration. In general, this path integral is ill-defined and needs to be equipped with an appropriate ultraviolet (UV) regularization, \emph{e.g.}, via an UV-cutoff $\Lambda$,
\be\label{eq:pireg}
Z\,\rightarrow\, Z_\Lambda \equiv \int \mathcal{D}_\Lambda g \, e^{-S_\Lambda[g]} \, . 
\ee
In continuum theories the key question is then, what are the conditions on $\mathcal{D}_\Lambda g$ and $S_\Lambda[g]$ so that the continuum limit $\Lambda \rightarrow \infty$,
\begin{equation}
    \lim_{\Lambda\to\infty}\mathcal{D}_\Lambda g \, e^{-S_\Lambda[g]}=\mathcal{D} g \, e^{-S^{\rm bare}[g]}\,,
\end{equation}
is well-defined.
This is in contrast to theories of gravity where the spacetime is fundamentally discrete. In this case $\Lambda$ acquires a physical interpretation as a ``hard cutoff'', set by the inverse of the minimal spacetime length, and UV regularization is trivial. Canonically, this comes at the expense of a fundamental breaking of Lorentz symmetry, which is strongly constrained by low-energy experiments and observations~\cite{HAWC:2019gui}. 
This conclusion may be avoided within 
theories where UV-regularity originates from a smooth cutoff or due to UV/infrared(IR) mixing effects~\cite{Crowther:2017pho}. 

The focus of our letter is on continuum approaches to quantum gravity, specifically including the asymptotic safety program~\cite{Percacci:2017fkn,Reuter:2019byg} and non-local ghost-free gravity \cite{Biswas:2005qr,Modesto:2011kw,Biswas:2011ar}. The two most promising technical tools in this context are the lattice techniques \cite{Ambjorn:2012jv,Laiho:2016nlp,Loll:2019rdj} based on dynamical triangulations and the analytical methods of the functional renormalization group (FRG) \cite{Wetterich:1992yh,Morris:1993qb,Reuter:1996cp}. 

Monte Carlo simulations within the Causal Dynamical Triangulation (CDT)~\cite{Ambjorn:2012jv,Loll:2019rdj} and Euclidean Dynamical Triangulation (EDT) \cite{Laiho:2016nlp} programs investigate the existence of a continuum limit utilizing the path integral~\eqref{eq:pi} as a starting point. In this case, $S_\Lambda[g]$ is taken as the Einstein-Hilbert action and the regularization is implemented by approximating the continuous spacetime by a piecewise linear geometry built from simplices. The edge-length of the discrete building blocks then provides the UV cutoff. One can then study the phase space of geometries as a function of the bare coupling constants entering $S_\Lambda[g]$. The goal of the program is then to identify a second order phase transition where the regulator can be removed by sending the number of building blocks to infinity and the edge length to zero. Evidence supporting the existence of such a phase transition has been found in spherical and also toroidal topologies~\cite{Ambjorn:2011cg,Ambjorn:2012ij,Ambjorn:2016mnn,Gizbert-Studnicki:2017yop,Laiho:2016nlp}, indicating that the continuum limit indeed exists.

An alternative approach to the problem is provided by the FRG adapted to gravity~\cite{Reuter:1996cp}. In this case one introduces a fixed but arbitrary background metric $\gb_{\mu\nu}$ and considers fluctuations in this spacetime.
The FRG implements the Wilsonian renormalization by integrating out these fluctuations ``shell-by-shell'' in momentum space. This produces a one-parameter family of effective actions $\Gamma_k$ labeled by an IR momentum cutoff $k$, such that $\Gamma_k$ incorporates all quantum fluctuations with momenta $p^2>k^2$. A well-defined quantum field theory is associated with a complete renormalization group (RG) trajectory connecting $\Gamma_{k = \infty}$ and $\Gamma_{k = 0}$. In particular, in the limit $k\to0$ all quantum fluctuations are integrated out and one recovers the textbook effective action associated with the path integral~\eqref{eq:pi}. In order to ensure that all physical observables remain finite in the UV, one requests that in the limit $k \rightarrow \infty$ the action $\Gamma_k$ approaches a fixed point, $\lim_{k \rightarrow \infty} \Gamma_k = \Gamma_*$, where the value of all dimensionless coupling constants is finite. 

The common root in terms of the path integral \eqref{eq:pi} suggests a relation between the fixed-point action $\Gamma_*$ and the bare action $S^{\rm bare}$. The precise map between these two objects is encoded in the  ``reconstruction formula''~\cite{Manrique:2008zw,Manrique:2009tj,Morris:2015oca}.  Solving this equation is known as the reconstruction problem. Its resolution is key to reconstruct the path integral from the FRG. While a full reconstruction is still out of reach, understanding the structures which can appear in $\Gamma_\ast$ and $S^{\mathrm{bare}}$ may already provide important insights to the problem. In its simplest approximation, the reconstruction formula takes the form of a one-loop integral, cf. eq.~\eqref{eq:reconstruction}.
On these grounds, one may expect that this relation gives rise to non-local terms, similarly to those encountered in a one-loop computation of the gravitational effective action in an effective field theory setting~\cite{Akhundov:1996jd,Bjerrum-Bohr:2013bxa}. Additionally, there is the possibility that either $\Gamma_\ast$ or $S^{\rm bare}$ take a very simple form which is then shadowed by the contributions of the reconstruction formula.
 
The goal of our letter is to evaluate the reconstruction formula using non-local heat-kernel techniques~\cite{Barvinsky:1993en,Codello:2012kq}, since this allows to track the appearance of non-localities and non-analytic terms. Concretely, we start from a local bare action, taken as the Einstein-Hilbert action, and check whether this is mapped onto a non-local fixed-point action. 
Our analysis identifies the presence or absence of fundamental dimensional scales as the  guiding principle in the computation of the relevant structures. We find that in the absence of a fundamental non-locality scale the continuum limit $\Lambda \rightarrow \infty$ leads to a rather simple relation between $\Gamma_\ast$ and $S^{\rm bare}$. In particular there are no non-localities generated up to quadratic order in the spacetime curvature. In particular, the fundamental scale invariance \cite{Wetterich:2019qzx,Wetterich:2020cxq} enjoyed by $\Gamma_\ast$ actually plays a crucial role in fixing the structure of the reconstruction map, as it prevents the occurrence of UV divergences in the dimensionless quantities appearing in the reconstruction map. Conversely, if scale invariance is broken by a fundamental mass scale, the reconstruction map is affected by logarithmic divergences. Finally, if the UV-scale $\Lambda$ has the interpretation of a physical UV-cutoff or if a fundamental non-locality scale exists the reconstruction map contains non-local terms.

%--------------------------------------------------
\section{The reconstruction problem}
%--------------------------------------------------

The gravitational path integral over all possible metric configurations~\eqref{eq:pi} is ill-defined, and one can think of different strategies to define a suitable regularization. The FRG provides a well-defined recipe to translate a functional integral into a functional integro-differential equation. In this approach, one introduces a fixed but arbitrary background $\bar{g}$ and considers arbitrary fluctuations $h$ on this background. The measure in the regularized path integral~\eqref{eq:pi} is complemented by an IR regulator term for the fluctuation fields
\be\label{eq:reg}
\Delta S_k[h]=\frac{1}{2}\int_x\,h \, \mathcal{R}_k \, h \,.
\ee
The function $\mathcal{R}_k$ provides a mass of order $k$ to fluctuations with momenta with $p^2 \lesssim k^2$ and decays sufficiently rapidly for momenta $p^2 \gtrsim k^2$. For $k \rightarrow \infty$, $\Delta S_k[h]$ gives an infinite mass to all fluctuations while for $k=0$ the regulator vanishes so that all quantum fluctuations are integrated out and one recovers the path integral~\eqref{eq:pireg}. Following standard textbook steps~\cite{Percacci:2017fkn,Reuter:2019byg}, one can then derive a flow equation---the Wetterich equation---for the effective average action~$\Gamma_k$ \cite{Wetterich:1992yh,Morris:1993qb},
\begin{equation}\label{eq:floweq}
    k \partial_k \Gamma_k =\frac{1}{2} \mathrm{STr} \left[\left( \Gamma_k^{(2)}+\mathcal{R}_k \right)^{-1} k\partial_k \mathcal{R}_k \right]\,.
\end{equation}
The flow of the effective average action is driven by the Wilsonian shell-by-shell integration of fluctuations $h$ with momenta $p^2\approx k^2$. 
Solutions of the flow equation are called RG trajectories $k \mapsto \Gamma_k$.
Trajectories that approach a fixed-point action $\Gamma_\ast$ as $k\to\infty$ can be safely extended to the UV. This feature ensures that there are no unphysical UV divergences in the construction.

The ``reconstruction problem''~\cite{Manrique:2008zw} is related to the quest of finding a relation between~$\Gamma_\ast$ and~$S^{\rm bare}$. More specifically, it boils down to the following question: what is the UV-regularized path integral corresponding to a given complete RG trajectory?  
While the knowledge of $\Gamma_k$, and in particular of its $k\to0$ limit, suffices to derive observables like scattering amplitudes, the knowledge of the bare action would provide a better way to compare different theories of quantum gravity, in particular those based on canonical quantization. 

Some insights into the formulation of the problem have been developed in~\cite{Manrique:2008zw,Manrique:2009tj,Morris:2015oca} and we review the main steps and results below. 
Let us consider a quantum field theory of a set of fields $\hat{\phi}$ described by a bare action~$S^{\rm bare}$. One can then construct a scale-dependent version of the generating functional that is also UV-regularized via, \emph{e.g.}, a sharp momentum cutoff $\Lambda$
\begin{equation}\label{eq:functintkL}
    e^{W_{k,\Lambda}}=\int\mathcal{D}_\Lambda \hat{\phi} \exp \left\{-S_\Lambda[\hat{\phi}]-\Delta S_k[\hat{\phi}]+\int d^4x\,J\cdot\hat{\phi} \right\}\,.
\end{equation}
The path integral measure can be thought as a sum over the Fourier modes of the different fields. In the case of a single scalar field with Fourier coefficients $\alpha_p$, it reads
\begin{equation}\label{eq:measure}
    \mathcal{D}_\Lambda \hat{\phi}=\prod_{|p|\in[0,\Lambda]}\int_{-\infty}^{+\infty} d\alpha_p M^{-\alpha}\,.
\end{equation}
Here $M$ is a constant with mass-dimension one which is required, with an appropriate power $\alpha$, to render the measure dimensionless. 

Based on this IR- and UV-regularized functional integral, one can introduce a UV-regularized version of the effective average action $\Gamma_{k,\Lambda}$. Following the standard steps~\cite{Percacci:2017fkn,Reuter:2019byg}, one derives that the change of $\Gamma_{k,\Lambda}$ resulting from integrating out an infinitesimal shell of quantum fluctuations with momentum~$k$ is controlled by the following modified Wetterich equation
\begin{equation}\label{eq:floweq-UVreg}
    k \partial_k \Gamma_{k,\Lambda} =\frac{1}{2} \mathrm{STr}_\Lambda \left[\left( \Gamma_{k,\Lambda}^{(2)}+\mathcal{R}_k \right)^{-1} k\partial_k \mathcal{R}_k \right]\,.
\end{equation}
Here $\STr_\Lambda[\hat{X}]\equiv\STr[\theta(\Lambda^2-p^2)\hat{X}]$ is the trace over momenta $p^2\leq\Lambda^2$. Owed to the fall-off conditions imposed on the regulator, the factor $k\partial_k \mathcal{R}_k$ also provides a UV-regularization of the trace. This allows to remove the UV-cutoff, sending $\Lambda \rightarrow \infty$, in a trivial way. This results in the standard Wetterich equation \eqref{eq:floweq}.

Formally, the relation between $S_\Lambda$ and $\Gamma_{\Lambda,\Lambda}\equiv\Gamma_\Lambda^\Lambda$ can be found by considering the generating functional for the effective average action $\Gamma_{k,\Lambda}$, derived from the regularized path integral. In the case of gravity it reads
\begin{equation}\label{eq:functintkL2}
\begin{split}
    e^{-\Gamma_{k,\Lambda}} = & \, \int \mathcal{D}_\Lambda h  \exp \Big( 
        -S_{\Lambda}[h] \\ & + \int_x 
        (h - \langle h \rangle) \fdv{\Gamma_{k,\Lambda}}{h}  
        -\Delta S_k[h - \langle h \rangle]
    \Big)  \,,
\end{split}
\end{equation}
where $\langle h \rangle$ denotes the expectation value of the field.
For $k \rightarrow \infty$ the regulator  essentially acts as a $\delta$-distribution, localizing the path integral on configurations $h - \langle h \rangle = 0$. 

Finally, to extract the $\Lambda$ dependence of $S_\Lambda$ and study the limit $\Lambda\to\infty$, one can set $k=\Lambda$ and expand  eq.~\eqref{eq:functintkL2} in powers of $\hbar$ (see~\cite{Manrique:2008zw} for details). At the one-loop level, the $\Lambda$-dependent Wilsonian and effective average actions are related by the following equation \cite{Manrique:2008zw}
\begin{equation}\label{eq:reconstruction}
	\Gamma^\Lambda_\Lambda = S_\Lambda + \frac{1}{2}\STr_\Lambda \left[ \ln \left( \left( S_\Lambda^{(2)} + \mathcal{R}_\Lambda  \right)
	{M}^{-2\alpha}  \right) \right]\,.
\end{equation}
In \cite{Manrique:2008zw,Manrique:2009tj,Morris:2015oca}, this relation has been studied in the approximation where $\Gamma_\Lambda^\Lambda$ and $S_\Lambda$ take the form of the Einstein-Hilbert action, i.e.,
\begin{equation}\label{eq.einstein}
    S_\Lambda=\frac{1}{16\pi \Breve{G}_\Lambda}\int d^4x \sqrt{g} \left(2\Breve{\lambda}_\Lambda-R\right)\, ,
\end{equation}
and similar for $\Gamma_\Lambda^\Lambda$. In this way, the works derived a relation between Newton's coupling and the cosmological constant at the bare and effective level. Furthermore,  \cite{Morris:2015oca} pointed out that one can resort to a special class of regulator functions $\mathcal{R}_\Lambda$, so-called compatible cutoff,  for which the trace in eq.\ \eqref{eq:reconstruction} takes a particularly simple form.

This provides the starting point of our analysis, which extends the previous construction to the inclusion of non-local terms. This extension is motivated by the following observation: A first inspection of the reconstruction formula suggests that a local bare action can be mapped onto a non-local fixed-point action. This would have important consequences, since it would indicate that the presence of non-localities in the fixed-point action $\Gamma_\ast$ does not necessarily entail a breakdown of locality in the bare theory. Indeed, while the effective action $\Gamma_0$ is expected to be non-local due to integrating out quantum fluctuations across all scales, it is an open question whether the theory build on $\Gamma_*$ is fundamentally local.

%------------------------------------------------------------------------------
\section{The reconstruction map is local}
%------------------------------------------------------------------------------
We now evaluate \eqref{eq:reconstruction} in an approximation which is capable of tracking the presence of non-local terms. For the sake of clarity we limit the present discussion to the key steps in the computation and the interpretation of our result, referring to App.\ \ref{App.A} for the technical details. 

For concreteness, we take $S_\Lambda$ to be a UV-cutoff-dependent version of the Einstein-Hilbert action, eq.~\eqref{eq.einstein}, supplemented by a gauge-fixing term implementing harmonic gauge and the corresponding ghost contributions, cf.\ eq.\ \eqref{eq.A2}. The first step in the evaluation then consists in computing the Hessian~$S_\Lambda^{(2)}$ in a fixed but arbitrary background. Decomposing the graviton fluctuations into their traceless and trace-part the result becomes block-diagonal in field space~\cite{Percacci:2017fkn,Reuter:2019byg}. The components of $S_\Lambda^{(2)}$ are linear functions of the Laplace operator $\square \equiv - \bar{g}^{\mu\nu} \bar{D}_\mu \bar{D}_\nu$ supplemented by an endomorphism $\mathbb{E}$.  In order to facilitate the computation, we chose the regulator $\cR_\Lambda$ to be of Type II  in the classification introduced in \cite{Codello:2008vh} and  perform the computation without specifying a cutoff function. This choice entails that all derivatives and curvature tensors contained in $S^{(2)}_\Lambda + \cR_k$ combine into differential operators of the form
\be\label{eq.defDelta}
\Delta\equiv\square+\mathbb{E}.
\ee
The trace on the right-hand side of eq.\ \eqref{eq:reconstruction} then turns in a sum of traces capturing the contributions of the fluctuation fields. Each contribution can be written as
\begin{equation}
  \Tr[W(\Delta)] = \int_{0}^{\infty} \dd{s} \widetilde{W}(s) \Tr\left[e^{-s \Delta}\right]\, ,
  \label{eq:wtrace}
\end{equation}
where $\widetilde{W}(s)$ is the inverse Laplace transform of $W(z)$. 
 For the Einstein-Hilbert action, the functions $W$ have the general form
\begin{equation}
W(z)=\theta(\Lambda^2-z)\ln \left( \alpha \left(\frac{z}{\Lambda^2}+r_\Lambda\left(\frac{z}{\Lambda^2}\right)+\beta\right)\right)\,,
\end{equation}
where $\alpha$ and $\beta$ are dimensionless constants and $r_\Lambda$ is the dimensionless version of the regulator~$\cR_\Lambda$, which is introduced in the Hessian through the replacement rule~\eqref{eq:replrule}.
The trace on the right-hand side of eq.~\eqref{eq:wtrace} is evaluated using non-local heat-kernel techniques~\cite{Barvinsky:1993en,Codello:2012kq}. Restricting to terms up to quadratic order in the curvature, the resulting general expression for the trace is given in eq.\ \eqref{non_local}. As indicated by its name, the non-local heat-kernel contains non-local functions of the Laplacian acting on spacetime curvatures, which are typically omitted in standard computations building on a derivative expansion. 

The relation between $S_\Lambda$ and $\Gamma_\Lambda^\Lambda$ is obtained by summing the trace-contributions of all fluctuation fields. Structurally, the result takes the form
\begin{equation}\label{eq:recons-form-factors}
    \begin{aligned}
    \Gamma^\Lambda_\Lambda - S_\Lambda &= \frac{1}{32 \pi^2} \int \dd[4]{x} \sqrt{g}
     \left( \mathcal{Q}_\Lambda + \mathcal{Q}_G R \right. \\ & \left. \,\, +  R\, \mathcal{Q}_R \left( \frac{\Box}{\Lambda^2} \right) R  + R_{\mu \nu}\, \mathcal{Q}_{Ric} \left( \frac{\Box}{\Lambda^2} \right)R^{\mu \nu}\right) \,.
\end{aligned}
\end{equation}
The general expressions for the functions $\mathcal{Q}_{i}$ are given in eqs.\ \eqref{eq.Qfct1} and \eqref{eq.Qfct2}, while their explicit forms for mass-type regulators are stated in eqs.\ \eqref{eq.Qfct1ev} and \eqref{eq.Qfct2ev}. These formulas constitute the main computational result of our work.

Based on this result, we make the following crucial observations:
\begin{enumerate}
    \item All functions $\cQ_i$ depend on the UV-cutoff $\Lambda$. In addition, $\cQ_R\left(\Box/\Lambda^2\right)$ and $\cQ_{Ric}\left(\Box/\Lambda^2\right)$ also depend on the Laplacian $\Box$. This dependence gives rise to non-local terms. Thus for $\Lambda$ being finite the relation between $S_\Lambda$ and $\Gamma^\Lambda_\Lambda$ is non-local.
    \item As indicated in eq.\ \eqref{eq:recons-form-factors} the dependence on $\Box$ appears in the combination $x \equiv \Box/\Lambda^2$ only. This feature is independent of the presence of other dimensionful scales as, e.g., $M$ or $\breve{G}_\Lambda$. The 
    $\Box$-dependence of $\cQ_R$ and $\cQ_{Ric}$ drops out in the continuum limit $\Lambda \rightarrow \infty$, cf.\ eq.\ \eqref{limitLambda}.
    This result is independent of the choice of regulator, as long as~$\cR_\Lambda$ satisfies the standard UV-behavior. Thus the relation between $S_\Lambda$ and $\Gamma^\Lambda_\Lambda$ is local in this limit. 
    \item Owed to their dimensional nature, the functions $\cQ_\Lambda$ and $\cQ_G$ scale as $\Lambda^4$ and $\Lambda^2$, respectively. In the presence of fundamental dimensionful quantities like $M$ and $\breve{G}_\Lambda$, we also encounter logarithmic divergences in $\cQ_R$ and $\cQ_{Ric}$ and in the dimensionless counterparts of $\cQ_\Lambda$ and $\cQ_G$. The structure of these divergences resembles the one encountered in the one-loop effective action~\cite{Satz:2010uu,Donoghue:2015nba,Donoghue:2017pgk,Percacci:2017fkn,Ohta:2020bsc}. 
    \item If the UV-completion is provided by an RG fixed point, fundamental scale invariance ensures that there is no other dimensionful scale than $\Lambda$, i.e., one must have $M\propto \Lambda$ and $\breve{G}_\Lambda \propto \Lambda^{-2}$ based on dimensional grounds. From eqs.\ \eqref{eq.Qfct1ev} and \eqref{eq.Qfct2ev} one then concludes that all logarithmic divergences vanish in this case. The remaining infinities are dictated by the mass-dimension of the operators. This implies that the dimensionless couplings in $S^{\rm bare}$ and $\Gamma_\ast$ differ by finite contributions. In contrast, if a fundamental mass scale exists in the theory, scale invariance is broken and all form factors display logarithmic divergences.
\end{enumerate}

The local nature of the reconstruction map in the continuum limit can also be understood intuitively based on a dimensional argument. In general, non-local form factors depending on a single momentum scale will be of the form $\cQ\left( \Box/M^2 \right)$. The only way to retain non-localities in the UV is by having a finite scale $M$ in the bare theory. Prototypical cases where this may be realized are
\begin{itemize}
    \item theories that are fundamentally discrete. In this case $M$ would be a finite physical cutoff $\Lambda_{\rm phys}$, possibly corresponding to a minimal length $l_{\rm min}\sim \Lambda_{\rm phys}^{-1}$. Then the form factors can retain a non-trivial momentum dependence, since the ``UV limit'' is $\Lambda\to\Lambda_{\rm phys}$.
    \item theories containing a fundamental non-locality scale $M$. In this case the bare theory contains a fundamental scale which, by definition, gives rise to the dimensionless ratios $(\Box/M^2)$. A prototypical example is non-local, ghost-free gravity \cite{Biswas:2005qr,Modesto:2011kw,Biswas:2011ar}.
\end{itemize}

If no such scale is present, all form factors are of the form $\cQ_i(p^2/\Lambda^2)$ and reduce to the constant $\cQ_i(0)$ for finite~$p^2$ and to $\cQ_i(1)$ in the continuum limit $p^2\to\Lambda^2\to\infty$. This enforces the locality of the form factors depending on a single momentum scale in the continuum limit. Notably, this argument does not assume a specific form of $S^{\rm bare}$. On this basis one expects that our conclusion about the locality of the reconstruction map is actually independent of the specific form of $S_{\Lambda}$ and holds beyond the specific example worked out in this letter.\footnote{Remarkably, the dimensional argument is not specific to the reconstruction map. It points to the intriguing conclusion that the existence of an RG fixed point requires the absence of non-localities in the quadratic part of $\lim_{\Lambda\to\infty}\Gamma_\Lambda^\Lambda$, as the latter would explicitly break scale invariance in the continuum limit (also see~\cite{Benedetti:2011ct} for a related argument). This also identifies a fundamental difference between asymptotically safe gravity and non-local ghost-free gravity: under the proviso that the non-locality scale does not drop out when applying the reconstruction map $S^{\rm bare} \mapsto \lim_{\Lambda \rightarrow \infty} \Gamma_\Lambda^\Lambda$, this additional dimensionful scale may prevent that the theory develops an RG fixed point controlling its UV-limit.}

Going to higher powers in the spacetime curvature one inevitably encounters form factors $\cQ_i$ depending on more than one momentum scale \cite{Knorr:2019atm,Pawlowski:2020qer}. In this case one can then form dimensionless ratios of the momenta. A simple example of where a non-trivial ratio of momenta appears, is a vertex $\left({\Box^{-1}} h \right) \left( \Box h \right) h$, which will be proportional to ${p_2^2}/{p_1^2}$ in momentum space. These ratios will survive also in the continuum limit. We stress that non-localities in interaction vertices have a quite different status than the one appearing in the two-point function, as the former can be compatible with general physics requirements \cite{tHooft:1973wag}. Analyzing this type of potential non-localities is beyond the scope of the present work though.

%------------------------------------------------------------
\section{Summary and Discussion}
%------------------------------------------------------------

Traditionally, the investigation of asymptotic safety in pure gravity and gravity-matter systems builds on solving the functional renormalization group equation for the effective average action $\Gamma_k$ \cite{Reuter:1996cp}. In this framework, the UV-completion of a theory is provided by a suitable UV-fixed point which appears as an outcome of the computation rather than an input. This raises the question about the relation between the fixed point action~$\Gamma_\ast$ and the bare action $S^{\rm bare}$ appearing in path-integral approaches to quantum gravity as the two do not necessarily coincide. The problem of re-obtaining the bare action from the functional renormalization group goes under the name of ``reconstruction problem''~\cite{Manrique:2008zw,Manrique:2009tj,Morris:2015oca}. In this letter, we investigated structural aspects of this problem. Specifically, the goal was to understand whether a local bare action could be mapped onto a non-local fixed-point action. In this light, we established the following results. Firstly, for a finite UV-cutoff $\Lambda$ the reconstruction map contains non-local terms. At the one-loop level worked out in this letter these structures agree with the ones found in the effective field theory treatment of gravity based on the Einstein-Hilbert action \cite{Satz:2010uu,Donoghue:2015nba,Donoghue:2017pgk,Percacci:2017fkn,Ohta:2020bsc}. Secondly, the non-local contributions drop out of the map once one takes the limit $\Lambda \rightarrow \infty$. Hence the continuum limit of the reconstruction map is local, at least at the level quadratic in the spacetime curvature.  This result holds in the absence of a physical discreteness scale or fundamental non-locality scale and implies that at quadratic order the bare and fixed-point actions are either both local or both non-local. Thirdly, in the absence of a fundamental scale the divergences in the continuum limit of the reconstruction map are strictly powerlaw and organized in such a way that the dimensionless coupling constants associated with powercounting relevant and marginal operators receive a finite corrections. In contrast, the presence of other fundamental mass scales makes the reconstruction map logarithmically divergent. 

Our results have deep implications when relating the Monte Carlo approach to quantum gravity based on Causal Dynamical Triangulations (CDT) \cite{Ambjorn:2012jv,Loll:2019rdj} and the functional renormalization group based on the effective average action. Since practical computations carried out in the CDT framework come with finite UV- and IR-cutoffs (which ought to be removed in the continuum limit though) our results establish that $\Gamma_\Lambda^\Lambda$ should actually contain non-local terms. This is in line with the explicit reconstruction of $\Gamma^\Lambda_k$ based on CDT correlation functions \cite{Knorr:2018kog} where non-local terms played a crucial role in reproducing the correlation functions obtained from the Monte Carlo simulations. For earlier results relating the spectral dimension obtained from the two approaches see \cite{Lauscher:2005qz,Reuter:2011ah,Reuter:2012xf}. 

At this stage, determining the momentum-dependent form factors in $\Gamma_\ast$ from first principles is still subject to ongoing research \cite{Christiansen:2015rva,Becker:2017tcx,Denz:2016qks,Bosma:2019aiu,Pawlowski:2020qer,Knorr:2021niv,Bonanno:2021squ,Fehre:2021eob}. We stress that the specific non-local structure of $\Gamma_\ast$  does not have direct implications for unitarity. Firstly, any physics implications should be checked at the level of the effective action $\lim_{k\rightarrow 0}\Gamma_k$ \cite{Knorr:2019atm,Bonanno:2020bil}. This applies specifically to the dressed propagator~\cite{Becker:2017tcx,Draper:2020bop,Platania:2020knd,Platania:2022gtt} and scattering amplitudes~\cite{Draper:2020bop,Draper:2020knh,Knorr:2022lzn}. Secondly, non-localities in the effective action could organize themselves in terms of the typical patterns encountered in string theory \cite{Alonso:2019ptb}. This would strengthen the connection the connection between asymptotic safety and string theory recently explored in~\cite{deAlwis:2019aud,Basile:2021euh,Basile:2021krk,Basile:2021krr,Gao:2022ojh,Ferrero:2022dpk} or non-local, ghost-free gravity discussed in~\cite{Knorr:2021iwv}. 

%----------------------------------------------
\begin{acknowledgments}
%----------------------------------------------
The authors would like to thank B. Knorr for interesting discussions. A.P. acknowledges support by Perimeter Institute for Theoretical Physics. Research at Perimeter Institute is supported in part by the Government of Canada through the Department of Innovation, Science and Economic Development and by the Province of Ontario through the Ministry of Colleges and Universities. A.P. is also grateful to the Radboud University for hospitality during various stages of this project.
\end{acknowledgments}

\let\clearpage\relax 
\onecolumngrid

\appendix

%---------------------------------------------------------
\section{Evaluating the reconstruction formula at one-loop}
\label{App.A}
%---------------------------------------------------------

Using the saddle point expansion to one-loop order, the relation between $\Gamma^\Lambda_\Lambda$ and $S_\Lambda$ is given by the reconstruction formula~\eqref{eq:reconstruction}. This also entails the relation between $S^{\rm bare}$ and $\Gamma_\ast$ in the limit $\Lambda\to\infty$. 
In this appendix we give the technical details leading to the result \eqref{eq:recons-form-factors}. 
We then review the computation of the regulated Hessians appearing in the argument of the logarithm in App.~\ref{App.A1} and subsequently proceed with the evaluation of the supertrace via non-local heat kernel techniques in App.~\ref{App.A2}. Throughout the computation we employ a generic regulator $\mathcal{R}_\Lambda$ and resort to a specific choice in the final evaluation of the form factors $\mathcal{Q}_i$ in eq.~\eqref{eq:recons-form-factors} only. Notably, similar computations at the level of the FRGE \eqref{eq:floweq} have already been carried out in \cite{Satz:2010uu,Ohta:2020bsc}.

%---------------------------------------------------------
\subsection{Constructing the Trace Arguments}
\label{App.A1}
%---------------------------------------------------------
The evaluation of \eqref{eq:reconstruction} utilizes the background field method, decomposing the Euclidean spacetime metric $g_{\mu\nu}$ into a fixed but arbitrary background metric $\gb_{\mu\nu}$ and fluctuations $h_{\mu\nu}$. We employ the linear split
\be
g_{\mu\nu} = \gb_{\mu\nu} + h_{\mu\nu} \, . 
\ee
It is then technically convenient to decompose the fluctuations into their trace and traceless parts
\be\label{tracedec}
h_{\mu\nu} = \hat{h}_{\mu\nu} + \frac{1}{4} \gb_{\mu\nu} h \, , \qquad {\rm with} \qquad \gb^{\mu\nu} \hat{h}_{\mu\nu} = 0 \, , \; h = \gb^{\mu\nu} h_{\mu\nu} \, . 
\ee
Throughout this appendix, we use the conventions that tensors and covariant derivatives constructed from $\gb_{\mu\nu}$ carry a bar.

The next step determines the second variation $S^{(2)}_\Lambda[h; \bar{g}]$ from the Einstein-Hilbert action supplemented by a background gauge fixing term
\begin{equation}\label{ehgaugefixed}
    S_\Lambda[h; \bar{g}] = \frac{1}{16 \pi \Breve{G}_\Lambda} \int \dd[4]{x} \sqrt{g} \left( 2 \Breve{\bar{\lambda}}_\Lambda - R \right)+ \frac{1}{32 \pi \Breve{G}_\Lambda} 
    \int \dd[4]{x} \sqrt{\bar{g}} \bar{g}^{\mu \nu} F_\mu \, F_\nu 
    \,.
\end{equation}
For concreteness, we adopt harmonic gauge
\be
F_\mu = \Db^\nu h_{\mu\nu} - \frac{1}{2} \Db_\mu h \, . 
\ee
This has the technical advantage that all derivatives appearing in $S_\Lambda^{(2)}$ combine into Laplacians $\Box = - \gb^{\mu\nu} \Db_\mu \Db_\nu$. The ghost action accompanying this choice is
\be\label{Sghost}
S^{\rm ghost}_\Lambda[C,\bar{C},h;\gb] = - \sqrt{2} \int \dd^4x \sqrt{\gb} \, \bar{C}_\mu \left[ \delta^\mu_\nu \Db^2 + \Rb^\mu{}_\nu \right] C^\nu \, ,
\ee
where we simplified the expression by restricting ourselves to the terms quadratic in the fluctuation fields.

We then expand \eqref{ehgaugefixed} in powers of the fluctuation field. The terms quadratic in $h_{\mu\nu}$ read
\begin{equation}\label{eq.A2}
    S^{\rm quad}_\Lambda[h ; \bar{g}] = \frac{1}{32 \pi \Breve{G}_\Lambda} \int \dd[4]{x} \sqrt{\bar{g}} \; h_{\mu \nu} \left(K\indices{^\mu^\nu_\rho_\sigma} \, \Box + 
    U\indices{^\mu ^\nu _\rho _\sigma} \right) h^{\rho \sigma} \,,
\end{equation}
with the tensors $K$ and $U$ given by
\begin{equation}\label{def.tensors}
\begin{aligned}
    & K\indices{^\mu^\nu_\rho_\sigma} = \frac{1}{4} \left( \delta^\mu_\rho \delta^\nu_\sigma + \delta^\mu_\sigma \delta^\nu_\rho -
    \bar{g}^{\mu \nu} \bar{g}_{\rho \sigma} \right) \,, \\
    & U\indices{^\mu^\nu_\rho_\sigma}  = K\indices{^\mu^\nu_\rho_\sigma} \left( \bar{R} - 2 \Breve{\bar{\lambda}}_\Lambda \right) 
    + \frac{1}{2} \left( \bar{g}^{\mu \nu}  \bar{R}_{\rho \sigma} + \bar{g}_{\rho \sigma} \bar{R}^{\mu \nu} \right)
    -\frac{1}{4} \left( \delta^\mu_\rho \bar{R}\indices{^\nu_\sigma} + \delta^\mu_\sigma \bar{R}\indices{^\nu_\rho} 
    + \delta^\nu_\rho \bar{R}\indices{^\mu_\sigma} + \delta^\nu_\sigma \bar{R}\indices{^\mu_\rho} \right)
    -\frac{1}{2} \left( \bar{R}\indices{^\nu_\rho^\mu_\sigma} + \bar{R}\indices{^\nu_\sigma^\mu_\rho} \right)\,.
\end{aligned}
\end{equation}
Upon substituting the decomposition \eqref{tracedec}, this expression simplifies to
\begin{equation}\label{Squad}
    S^{\rm quad}_\Lambda[h ; \Bar{g}] = \frac{1}{32 \pi
    \Breve{G}_{\Lambda}} \int \dd[4]{x} \sqrt{\Bar{g}} \left( 
    \frac{1}{2} \hat{h}_{\mu \nu} \left( \Box
    -2\Breve{\bar{\lambda}}_{\Lambda}+\bar{R} \right) \hat{h}^{\mu
    \nu}- \frac{1}{8} h \left( \Box -
    2\Breve{\bar{\lambda}}_{\Lambda} \right) h - \bar{R}_{\mu
    \nu}
    \hat{h}^{\nu \rho} \hat{h}\indices{^\mu_\rho} - \bar{R}_{\rho \sigma \mu \nu} \hat{h}^{\rho \nu} \hat{h}^{\sigma \mu} \right).
\end{equation}
Based on eqs.\ \eqref{Squad} and \eqref{Sghost} it is then straightforward to read off the Hessians
\begin{equation}
\begin{aligned}\label{eq:hessian}
    &(32 \pi \Breve{G}_\Lambda) \, (S^{(2)}_{\rm T})\indices{^\gamma^\sigma_\alpha_\beta} = \left(\Box -2 \Breve{\bar{\lambda}}_{\lambda}\right) \Pi\indices{^\gamma^\sigma_\alpha_\beta} + \frac{2}{3}\bar{R}\, \Pi\indices{^\gamma^\sigma_\alpha_\beta} - 
    \bar{C}\indices{^\gamma_\alpha^\sigma_\beta} -
    \bar{C}\indices{^\gamma_\beta^\sigma_\alpha}\,,\\
    & (32 \pi \Breve{G}_\Lambda) S^{(2)}_{\rm S} = - \frac{1}{4} \left( \Box - 2 \Breve{\bar{\lambda}}_{\lambda} \right)\,, \\
    & (S^{(2)}_{\rm V})^\mu{}_\nu =  \sqrt{2} \left( \Box \, \delta^\mu_\nu - \Rb^\mu{}_\nu \right) \, . 
\end{aligned}
\end{equation}
Here T, S, V label the contribution in the traceless-tensor sector, scalar fluctuations and ghosts, respectively. Moreover, we denote the unit matrix on the space of traceless tensors by $\Pi\indices{^\mu^\nu_\rho_\sigma} = \frac{1}{2} \left( \delta^\mu_\rho \delta^\nu_\sigma + \delta^\nu_\rho \delta^\mu_\sigma \right) -\frac{1}{4} \bar{g}_{\rho \sigma} \bar{g}^{\mu \nu}$.
Comparing the expressions \eqref{eq:hessian} to the definition \eqref{eq.defDelta} yields the explicit expression for the endomorphism $\mathbb{E}$ in each sector
\begin{equation}
    (\mathbb{E}_{\rm T})\indices{^\gamma^\sigma _\alpha _\beta} = \frac{2}{3} \, \bar{R} \, \Pi\indices{^\gamma^\sigma_\alpha_\beta} - 
    \bar{C}\indices{^\gamma_\alpha^\sigma_\beta} -
    \bar{C}\indices{^\gamma_\beta^\sigma_\alpha} \,,\qquad \mathbb{E}_{\rm S} = 0 \, , \qquad 
    (\mathbb{E}_{\rm V})\indices{^\mu_\nu} = -\bar{R}\indices{^\mu_\nu}\,.
\end{equation}
Combining these endomorphisms with the $\Box$-operator then yields the differential operators appearing inside the trace
\be\label{diffops}
    \Delta_{\rm T} \equiv \Box + \mathbb{E}_{\rm T} \,, \qquad
    \Delta_{\rm S} \equiv \Box\,, \qquad
    \Delta_{\rm V} \equiv \Box + \mathbb{E}_{\rm V}\, . 
\ee
As final ingredient, we have to specify the regulator function $\cR_\Lambda$. The corresponding functions are generated by the replacement rule
\be\label{eq:replrule}
\Delta_i \mapsto \Delta_i + \Lambda^2 \, r_\Lambda(\Delta_i/\Lambda^2) \, , \qquad i = \{ {\rm T,S,V} \} \, , 
\ee
where $r_\Lambda(z)$ is a dimensionless profile function. This rule fixes $\mathcal{R}_\Lambda$ in terms of $r_\Lambda(z)$, including all prefactors. Notably,  the regulator also contains the endomorphism piece, which in the nomenclature introduced in \cite{Codello:2008vh}, corresponds to a regulator of type II. This choice aids the subsequent computation since the functions appearing within the trace become simple in the sense that they depend on a single type of differential operator only.
%---------------------------------------------------------
\subsection{Evaluation of the trace via non-local heat kernel techniques}
\label{App.A2}
%---------------------------------------------------------
At this stage, we have all the ingredients entering into the trace appearing on the right-hand side of eq.\ \eqref{eq:reconstruction}. Using that $S_\Lambda^{(2)}$ and $\cR_\Lambda$ are block diagonal and utilizing the definitions \eqref{diffops}, we obtain
\be\label{eq.t1}
\begin{split}
T = & \, \frac{1}{2} \Tr^{\rm T}_\Lambda \left[ \ln\left( \frac{1}{32 \pi \Breve{G}_\Lambda M^4} \left( \Delta_{\rm T} + \Lambda^2 r_\Lambda \left(  \frac{\Delta_{\rm T}}{\Lambda^2}\right) - 2\Breve{\bar{\lambda}}_\Lambda  \right) \right) \right]  \\ & \, 
    +\frac{1}{2} \Tr^{\rm S}_\Lambda \left[ \ln \left( \frac{1}{128 \pi \Breve{G}_\Lambda M^4} \left( \Delta_{\rm S} + \Lambda^2 r_\Lambda \left(  \frac{\Delta_{\rm S}}{\Lambda^2}\right) - 2\Breve{\bar{\lambda}}_\Lambda \right)\right)\right] 
    -\Tr^{\rm V}_\Lambda \left[ \ln \left(\frac{1}{M^2} \left( \Delta_{\rm V} +\Lambda^2 r_\Lambda\left(\frac{\Delta_{\rm V}}{\Lambda^2}\right) \right) \right) \right]\, .
\end{split}
\ee
Here the superscripts $\{$T,S,V$\}$ indicate that the trace is over traceless symmetric tensors, scalars, and vectors, respectively. The subscript $\Lambda$ implies that fluctuations with eigenvalues $p^2 > \Lambda^2$ are excluded from the trace, i.e., $\Tr_\Lambda[\hat{X}]\equiv\Tr[\theta(\Lambda^2-p^2)\hat{X}]$.
Furthermore, we dropped a field-independent infinite constant. The structure of the trace arguments suggests writing eq.\ \eqref{eq.t1} in a unifying way,
\be\label{3traces}
T = \frac{1}{2} \Tr^{\rm T} \left[W^{\rm T} \left( \Delta_{\rm T} \right) \right] + \frac{1}{2} \Tr^{\rm S} \left[W^{\rm S} \left( \Delta_{\rm S} \right) \right] - \Tr^{\rm V} \left[W^{\rm V} \left( \Delta_{\rm V} \right) \right]\,,
\ee
with the function inside the trace having the general form
\be\label{eq.structure}
W^{i} (z) = \theta(\Lambda^2-z) \ln \left( \alpha_{i} \left( \frac{z}{\Lambda^2} + r_\Lambda \left( \frac{z}{\Lambda^2} \right)+ \beta_{i} \right) \right)\,.
\ee
The dimensionless constants can be read off from comparing eqs.\ \eqref{eq.t1} and \eqref{eq.structure} and read
\be
\begin{array}{lll}
 \alpha_{\rm T} = \frac{ \Lambda^4 }{32 \pi g_\Lambda M^4} \, , \qquad & 
 \alpha_{\rm S} = \frac{ \Lambda^4 }{128 \pi g_\Lambda M^4} \, , \qquad & 
 \alpha_V = \frac{\Lambda^2}{M^2} \, , \\[2ex]
 \beta_{\rm T} = -2 \lambda_\Lambda \,, & 
 \beta_{\rm S} = -2 \lambda_\Lambda \, , &
 \beta_{\rm V} = 0 \, . 
\end{array}
\ee
Here $g_\Lambda \equiv 
\Breve{G}_\Lambda \Lambda^2$ and $\lambda_\Lambda \equiv  \Breve{\bar{\lambda}}_\Lambda \Lambda^{-2}$ are the dimensionless Newton coupling and cosmological constant. At this point the evaluation of $T$ has been reduced to evaluating a trace of the form
\begin{equation}\label{eq:heat-kernel}
    \Tr \left[ W(\Delta) \right] = \int_0^\infty \dd{s} \widetilde{W}(s) \Tr \left[ e^{-s \Delta} \right]\,.
\end{equation}
Here $\widetilde{W}(s)$ is the inverse Laplace transform of $W(z)$ which, in the present computation, takes the generic form \eqref{eq.structure}.

 The trace of the heat kernel has the following expansion to second order in the curvature \cite{Barvinsky:1990up,Codello:2012kq}
\begin{multline}\label{non_local}
    \Tr \left[ e^{ -s \Delta } \right] = \frac{1}{ (4 \pi s )^2} \int \dd[4]{x} \sqrt{g} \tr \Big\{ \mathbf{1} - s \mathbb{E} + s \frac{R}{6} \mathbb{1} 
    + s^2 \Big( R_{\mu \nu} f_{Ric}(s\Box) R^{\mu \nu} \mathbb{1} + R f_{R}(s\Box) R\, \mathbb{1}  \\
    + R f_{RE} (s \Box) \mathbb{E} + \mathbb{E} f_E (s \Box) \mathbb{E} + \Omega_{\mu \nu} f_{\Omega}(s \Box) \Omega^{\mu \nu} \Big) \Big\}\,.
\end{multline}
Here $\mathbb{1}$ is the unit operator, $R$ and $R_{\mu \nu}$ are the Ricci scalar and Ricci tensor constructed out of $\bar{g}_{\mu \nu}$, and $\Omega_{\mu \nu}$ is the commutator of the covariant derivative, $\Omega_{\mu \nu} \equiv [D_{\mu} , D_{\nu}]$. The form factors $f_i(s\Box)$ are given by
\begin{align}\label{eq:ffunctt}
    f_{Ric}(x) & = \frac{f(x)-1 + \frac{x}{6}}{x^2} \,,  &   f_R (x) & = \frac{f(x)}{32} + \frac{f(x)-1}{8x} - \frac{f(x)-1 + \frac{x}{6}}{8x^2} \,,  &&& \\ \nonumber
    f_{RE} (x) & = -\frac{f(x)}{4} - \frac{f(x)-1}{2x}\,, \qquad
    & 
    f_{E} (x) & = \frac{f(x)}{2}\,, \qquad  &
    f_{\Omega} & = - \frac{f(x)-1}{2x}\,,
\end{align}
with the universal heat-kernel function $f$ given by
\begin{equation}
    f(x) = \int_0^1 \dd{\xi} e^{-x \xi (1 - \xi)}\,.
\end{equation}
The functions $f_{i}(x)$ are non-polynomial in their argument, so that the expansion \eqref{non_local} is non-local. Note that, in principle there could be a form factor associated with the square of the Riemann tensor. The Bianchi identity allows to map these contributions to the form factors $f_{Ric}(x)$ and $f_R(x)$ through the identity \cite{Knorr:2019atm} 
\be\label{mapff}
\int d^4x \sqrt{g} \left\{R^{\rho\sigma\mu\nu} \, \Box^n \,  R_{\rho\sigma\mu\nu} - 4 R^{\mu\nu} \, \Box^n \,  R_{\mu\nu} + R \, \Box^n \,  R \right\} = O(R^3) \, , \quad n \ge 1 \, . 
\ee
We then use this identity to eliminate the Riemann-squared terms, writing our results in the ``Ricci-basis'' spanned by the square of the Ricci scalar and Ricci tensor.

Substituting the non-local heat-kernel expansion \eqref{non_local} into  \eqref{eq:heat-kernel} leads to the following expression 
\begin{equation}
\begin{aligned}
\Tr \left[ W (\Delta) \right]  & = \frac{1}{ (4 \pi )^2} \int \dd[4]{x} \sqrt{g} \int_0^\infty \dd{s} \widetilde{W}(s) 
\bigg(  s^{-2}  \tr\{\mathbb{1}\}
    - s^{-1} \tr\{\mathbb{E}\} - \frac{R}{6}  s^{-1}  \tr\{\mathbb{1}\}    + R_{\mu \nu}  f_{Ric}(s\Box) R^{\mu \nu} \tr\{\mathbb{1}\} \\
    &
    + R   f_R(s\Box) R \tr\{\mathbb{1}\} 
    + R   f_{RE}(s\Box) \tr\{\mathbb{E}\} 
    +\tr \left\{\mathbb{E}   f_E(s\Box) \mathbb{E} \right\}
    + \tr \left\{\Omega_{\mu \nu}   f_\Omega(s\Box) \Omega^{\mu \nu} \right\} \bigg)\,.
\end{aligned}
\end{equation}
At this stage, the evaluation of the reconstruction formula has boiled down to evaluating the $s$-integrals and the traces over the internal space. With respect to the former, we note that the functions~\eqref{eq:ffunctt} are linear combinations of $f(x)$, $\frac{f(x) - 1}{x}$,
and $\frac{f(x) -1 + \frac{x}{6}}{x^{2}}$. The integrals including the inverse Laplace transform can then be expressed as integrals including the initial function $W(z)$ through the standard Mellin transform \cite{Codello:2010mj,Reuter:2019byg}
\be\label{Qfcts}
Q_n[W] = \int_0^\infty \dd{s} s^{-n} \widetilde{W}(s) = \frac{1}{\Gamma(n)} \int_0^\infty \dd{z} z^{n-1} W(z) \, , 
\ee
and its extension to form factors \cite{Knorr:2021niv}
\be\label{eq:master_integrals}
\begin{aligned}
    \mathcal{C}_1[W](z) &\equiv \int_0^\infty \dd{s} \widetilde{W}(s) f(sz) = 2 \int_0^{1/4} \dd{u} \frac{W(uz)}{\sqrt{1-4u}} \,, \\
    \mathcal{C}_2[W](z) &\equiv \int_0^\infty \dd{s} \widetilde{W}(s) \frac{f(sz) - 1}{sz} = - \int_0^{1/4} \dd{u} \sqrt{1-4u} \; W(uz) \,, \\
    \mathcal{C}_3[W](z) &\equiv \int_0^\infty \dd{s} \widetilde{W}(s) \frac{f(sz) - 1 + \frac{sz}{6}}{(sz)^2} =  \frac{1}{6} \int_0^{1/4} \dd{u} (1-4u)^{3/2} \, W(uz) \,.
\end{aligned}
\ee
The values of the traces over internal indices depend on the internal space. In the scalar sector $\Omega_{\mu\nu}=0$ and $\mathbb{E}^{\rm S} = 0$. Thus the only non-vanishing trace in this sector is
\be\label{tracescalar}
\tr^{\rm S}\{\mathbb{1}\} = 1 \, . 
\ee
For vectors and traceless symmetric tensors we encounter four distinct trace structures which evaluate to
\be\label{tracevector}
\begin{split}
& \, \tr^{\rm V}\{\mathbb{1}\} = 4 \, , \;\;
\tr^{\rm V}\{\mathbb{E}^{\rm V}\} = -R \, , \;\;
\tr^{\rm V}\{\mathbb{E}^{\rm V} \mathcal{C}(\Box) \mathbb{E}^{\rm V} \} = R_{\mu \nu} \mathcal{C} (\Box) R^{\mu \nu} \, , \; \;
\tr^{\rm V}\{\Omega_{\mu \nu} \mathcal{C}(\Box) \Omega^{\mu \nu}\} = -R_{\rho \sigma \mu \nu} \mathcal{C}(\Box) R^{\rho \sigma \mu \nu} \,,
\end{split}
\ee
and
\be\label{tracetensor}
\begin{split}
& \, \tr^{\rm T}\{\mathbb{1}\} = 9 \, , \qquad
\tr^{\rm T}\{\mathbb{E}^{\rm T}\} = 6R \, , \qquad \\[1.2ex] & \, 
\tr^{\rm T}\{\mathbb{E}^{\rm T} \mathcal{C}(\Box) \mathbb{E}^{\rm T}\} =  5 R \mathcal{C}(\Box) R 
    - 6 R_{\mu \nu} \mathcal{C} (\Box) R^{\mu \nu} + 3 R_{\mu \nu \rho \sigma} \mathcal{C}(\Box) R^{\mu \nu \rho \sigma} \, , \\[1.2ex]
& \, \tr^{\rm T}\{\Omega_{\mu \nu} \mathcal{C}(\Box) \Omega^{\mu \nu}\} = -6 R_{\rho \sigma \mu \nu} \mathcal{C}(\Box) R^{\rho \sigma \mu \nu} \,.
\end{split}
\ee
The traces in the vector and tensor sector still contain contributions which are not adapted to the Ricci-basis. We then rewrite the terms including the square of the Riemann tensor by means of \eqref{mapff}. For vectors, this implies
\be\label{vecmod}
\tr^{\rm V}\{\Omega_{\mu \nu} \mathcal{C}(\Box) \Omega^{\mu \nu}\} \simeq -4 R_{\mu \nu} \mathcal{C}(\Box) R^{\mu \nu}
+R \mathcal{C}(\Box) R \, , 
\ee
while in the tensor sector we obtain
\be
\begin{split}
& \tr^{\rm T}\{\mathbb{E}^{\rm T} \mathcal{C}(\Box) \mathbb{E}^{\rm T}\} \simeq  2 R \mathcal{C}(\Box) R 
    + 6 R_{\mu \nu} \mathcal{C} (\Box) R^{\mu \nu} \, ,   \\[1.2ex]
& \, \tr^{\rm T}\{\Omega_{\mu \nu} \mathcal{C}(\Box) \Omega^{\mu \nu}\}  \simeq -24 R_{\mu \nu} \mathcal{C}(\Box) R^{\mu \nu}
+6 R \mathcal{C}(\Box) R \, . 
\end{split}
\ee
In contrast to the results \eqref{tracevector} and \eqref{tracetensor}, these expressions are not exact but valid up to terms of third order in the spacetime curvature only.

At this point we can evaluate the traces in \eqref{3traces} explicitly in terms of the $Q$-functionals \eqref{Qfcts} and $\mathcal{C}$-functionals~\eqref{eq:master_integrals}. Utilizing the explicit result for the traces and limiting to terms quadratic in the spacetime curvature we obtain
\be
     \Tr^{\rm S}\left[W^{\rm S}(\Delta) \right] =  \frac{1}{16 \pi^2} \int \dd[4]{x} \sqrt{g} \left( Q_2 - \frac{Q_1}{6}R 
    +R_{\mu \nu} \mathcal{C}_3  R^{\mu \nu}
    + R 
    \left( \frac{1}{32} \mathcal{C}_1   
    + \frac{1}{8} \mathcal{C}_2
    - \frac{1}{8} \mathcal{C}_3 \right) R
    \right)\,, \\ 
\ee
as well as
\be
\Tr^{\rm V}[W^{\rm V}(\Delta)] =  \, \frac{1}{16 \pi^2} \int \dd[4]{x} \sqrt{g} \left( 4 Q_2 + \frac{Q_1}{3} R
    + R_{\mu \nu} \left(\frac{1}{2} \mathcal{C}_1 + 2 \mathcal{C}_2+  4 \mathcal{C}_3  \right)  R^{\mu \nu} 
     + R \left( \frac{3}{8} \mathcal{C}_1  
    + \frac{1}{2} \mathcal{C}_2  
    -\frac{1}{2} \mathcal{C}_3 \right) R \right)\,,    
\ee
and
\be
\Tr^{\rm T}[W^{\rm T}(\Delta)] = \frac{1}{16 \pi^2} \int \dd[4]{x} \sqrt{g} \left(9Q_2 -\frac{15Q_1 }{2}R
+ R_{\mu\nu} \left( 3 \mathcal{C}_1 + 12 \mathcal{C}_2 + 9 \mathcal{C}_3 \right) R^{\mu\nu} 
- R \left( 
\frac{7}{32} \mathcal{C}_1 + \frac{39}{8} \mathcal{C}_2 + \frac{9}{8} \mathcal{C}_3
\right) R 
\right) \, . 
\ee
All $\mathcal{C}$-functions are evaluated at $\mathcal{C}[W^i]\left(\Box\right)$ and we have suppressed their arguments for the sake of readability. The final result for the evaluated reconstruction formula is then obtained by substituting these traces into \eqref{3traces}
\begin{equation}\label{reconstruction-eval}
\begin{aligned}
    \Gamma^\Lambda_\Lambda - S_\Lambda=
    \frac{1}{32 \pi^2} & \int \dd[4]{x} \sqrt{g}
     \left( \mathcal{Q}_\Lambda + \mathcal{Q}_G R
     +  R\, \mathcal{Q}_R \left( \Box \right) R
     + R_{\mu \nu}\, \mathcal{Q}_{Ric} \left( \Box \right)R^{\mu \nu}
     \right)\,,
\end{aligned}
\end{equation}
The $\mathcal{Q}$-functionals appearing in the local part of the action are
\be\label{eq.Qfct1}
\begin{aligned}
    \mathcal{Q}_{\Lambda} = Q_2\left[W^{\rm S}\right] - 8 Q_2\left[W^{\rm V}\right] + 9 Q_2\left[W^{\rm T}\right] \,, \qquad
    \mathcal{Q}_{G} = - \frac{1}{6} Q_1\left[W^{\rm S}\right] - \frac{2}{3} Q_1\left[W^{\rm V}\right] - \frac{15}{2} Q_1\left[W^{\rm T}\right] \,. 
\end{aligned}
\ee
Furthermore, we have two non-local form factors appearing at the second order of the spacetime curvature:
\be\label{eq.Qfct2}
\begin{split}
    \mathcal{Q}_{R} = & \,  \frac{1}{32} \mathcal{C}_1\left[W^{\rm S}\right] + \frac{1}{8} \mathcal{C}_2\left[W^{\rm S}\right]
    - \frac{1}{8} \mathcal{C}_3\left[W^{\rm S}\right] - \frac{3}{4} \mathcal{C}_1\left[W^{\rm V}\right] 
    - \mathcal{C}_2\left[W^{\rm V}\right] + \mathcal{C}_3\left[W^{\rm V}\right]
    \\ & \;
    - \frac{7}{32} \mathcal{C}_1\left[W^{\rm T}\right] - \frac{39}{8} \mathcal{C}_2\left[W^{\rm T}\right]
    - \frac{9}{8} \mathcal{C}_3\left[W^{\rm T}\right]\,, \\
    \mathcal{Q}_{Ric} = & \, \mathcal{C}_3\left[W^{\rm S}\right] - \mathcal{C}_1\left[W^{\rm V}\right]-4\mathcal{C}_2\left[W^{\rm V}\right] - 8 \mathcal{C}_3\left[W^{\rm V}\right]+ 3 \mathcal{C}_1\left[W^{\rm T}\right] + 12 \mathcal{C}_{2}\left[W^{\rm T}\right]
    + 9 \mathcal{C}_3\left[W^{\rm T}\right]  \,. %\\
\end{split}
\ee
Eqs.\ \eqref{reconstruction-eval}-\eqref{eq.Qfct2} are the main result of our work. They give the relation between $S_\Lambda$ and $\Gamma_\Lambda^\Lambda$ tracking all non-local terms appearing at second order of the spacetime curvature.
%-----------------------------------------------------------
\subsection{Evaluating the Threshold Integrals}
%-----------------------------------------------------------
In order to draw conclusions about the non-local nature of \eqref{reconstruction-eval} one has to investigate the structure of the $Q$-functionals and form factors $\mathcal{C}_n$ and determine their limiting behavior as $\Lambda \rightarrow \infty$. In order to illustrate the generic behavior, we will resort to a mass-type regulator where $r_\Lambda\left(\Box/\Lambda^2\right) = 1$. This choice is not crucial for the restoration of locality in the limit $\lim_{\Lambda \rightarrow \infty} \Gamma_\Lambda^\Lambda$. Since all admissible regulators $\cR_\Lambda$ diverge as $k\rightarrow \Lambda \rightarrow \infty$, they all fulfill the compatibility condition \cite{Morris:2015oca}. We also checked by explicit computation that a Litim-type regulator leads to results that are qualitatively identical to the ones reported for $r_\Lambda = 1$.

When writing down the results for the threshold integrals, it is convenient to introduce the dimensionless variable
\be\label{xdef}
x \equiv \frac{\Box}{\Lambda^2} \, , \qquad x \le 1 \, , 
\ee
where we use that the theory does not contain fluctuations with eigenvalues of $\Box$ larger than $\Lambda^2$. 
Substituting \eqref{eq.structure} into \eqref{Qfcts} and evaluating the final integral then yields
\be\label{eq.Qeval}
\begin{split}
Q_2[W^i] = & \, \frac{\Lambda^4}{2}\Big(\ln(\alpha_i)  + \frac{1}{2} + \beta_i + (1+\beta_i)^2 \ln(1+\beta_i) - \beta_i (2+\beta_i)  \ln(2+\beta_i) \Big) \, ,  \\
Q_1[W^i] = & \, \Lambda^2 \Big( \ln(\alpha_i) + (2 + \beta_i)\ln(2+\beta_i) - (1+\beta_i)\ln(1+\beta_i) - 1\Big)\, . \\
\end{split}
\ee
Analogously the evaluation of \eqref{eq:master_integrals} gives
\be\label{eq.Ceval}
\begin{split}
\cC_1[W^i] = & \ln(\alpha_i) + \ln(1+\beta_i) -2 - \frac{1}{\tilde{x}} \ln\left(\frac{1-\tilde{x}}{1+\tilde{x}} \right) \, ,
 \\
\cC_2[W^i] = & -\frac{1}{6} \left( \ln(\alpha_i) + \ln(1+\beta_i) - \frac{8}{3} + \frac{8(1+\beta_i)}{x} - \frac{1}{\tilde{x}^3} \ln\left( \frac{1-\tilde{x}}{1 + \tilde{x}} \right) \right) \, , \\
\cC_3[W^i] = & \frac{1}{60} \left( \ln(\alpha_i) + \ln(1+\beta_i) - \frac{2}{5} - \frac{8(3(1+\beta_i)+x)(4(1+\beta_i) + x)}{3 x^2} - \frac{1}{\tilde{x}^5} \ln\left( \frac{1-\tilde{x}}{1 + \tilde{x}} \right)
\right) \, , 
\end{split}
\ee
where we introduced the combination $\tilde{x} \equiv \left( \frac{x}{4(1+\beta_i) + x } \right)^{1/2}$ to write the arguments of the logarithms in compact form.

At this stage we make the following observations. Firstly, the $Q$-functionals are independent of $x$. Hence these contributions are local. The non-trivial dependence on $\Box$ appears in the form factors $\cC_n$. A priori these expressions seems to contain non-localities in the form of inverse powers of the Laplacian, leading to poles as $x \rightarrow 0$. The computation of the residues associated with these negative powers reveals that there are also contributions from the logarithm-terms so that all negative powers of $x$ actually vanish. Consequently, the limit $x \rightarrow 0$ is finite. Explicitly, the expansion of \eqref{eq.Ceval} for $\Lambda\to\infty$ (corresponding to sending $x\to0$) yields
\begin{equation}\label{limitLambda}
\begin{aligned}
 \lim\limits_{\Lambda \rightarrow \infty} \mathcal{C}_1[W^{i}]  = \ln \left(\alpha_{i}(1+\beta_{i}) \right) \,, \quad
 \lim\limits_{\Lambda \rightarrow \infty} \mathcal{C}_2[W^{i}] = 
 -\frac{1}{6} \ln \left(\alpha_{i}(1+\beta_{i}) \right) \,, \quad
 \lim\limits_{\Lambda \rightarrow \infty} \mathcal{C}_3[W^{i}] = 
 \frac{1}{60} \ln \left(\alpha_{i}(1+\beta_{i}) \right)\,.
\end{aligned}
\end{equation}
The remarkable property of this result is that it is actually \emph{independent of the choice of regulator} as long as $\lim_{\Lambda \rightarrow \infty} r_\Lambda\left(\Box/\Lambda^2 \right) = 1$. This is readily verified by taking the limit $\Lambda \rightarrow \infty$ before evaluating the form factor integrals~\eqref{eq:master_integrals}.

For completeness, we give the explicit expression for the $\mathcal{Q}$-functionals appearing in \eqref{reconstruction-eval}. Substituting the results~\eqref{eq.Qeval} and \eqref{eq.Ceval} into~\eqref{eq.Qfct1} gives
\be\label{eq.Qfct1ev}
\begin{split}
\mathcal{Q}_\Lambda = & \, \frac{\Lambda^4}{2} \Big( 1- 20 \lambda + 10 (1 - 2 \lambda )^2 \ln(1 - 2 \lambda) + 40 \lambda (1-\lambda)  \ln(2-2\lambda) +12 \ln\left(\frac{\Lambda^2}{M^2} \right) 
- 10 \ln (32 \pi \breve{G}_\Lambda \Lambda^2) - 2 \ln(2)
 \Big) \, , \\
\mathcal{Q}_G = & \, 
\frac{\Lambda^2}{6} \Big( 50 
+ 46 (1-2\lambda) \ln(1-2\lambda)
-92 (1-\lambda) \ln(2-2\lambda)
- 96 \ln\left(\frac{\Lambda^2}{M^2} \right) 
+46 \ln ( 32 \pi \breve{G}_\Lambda \Lambda^2 )  - 6 \ln(2)
\Big)
\, , 
\end{split}
\ee
while the evaluation of yields \eqref{eq.Qfct2}
\be\label{eq.Qfct2ev}
\begin{split}
\mathcal{Q}_{R} = & \, -\frac{2627}{3600} + \frac{1417-2466 \lambda}{180 \, x} + \frac{1-36 \lambda(1-\lambda)}{15 \, x^2}  + \frac{97}{160} \ln(1-2\lambda) - \frac{1+ 10 \xt_0^2 -45 \xt_0^4}{60 \, \xt_0^5} \ln\left( \frac{1 - \xt_0}{1+ \xt_0} \right)  \\
& \,
+\frac{3-130 \tilde{x}_{\lambda}^2 + 30 \tilde{x}_{\lambda}^4}{160 x_{\lambda}^5} \ln\left( \frac{1 - \xt_\lambda}{1+ \xt_\lambda} \right) + \frac{31}{48} \ln\left(\frac{\Lambda^2}{M^2} \right) - \frac{97}{160} \ln ( 32 \pi \breve{G}_\Lambda \Lambda^2 )  - \frac{1}{16} \ln(2) \, , \\
\mathcal{Q}_{Ric} = & \, -\frac{41}{75} - \frac{4(127-430 \lambda)}{45 \, x} - \frac{16(1-20 \lambda(1-\lambda))}{15 \, x^2}  + \frac{7}{6} \ln(1-2\lambda)  + \frac{2 - 10 \xt_0^2 - 15 \xt_0^4}{15 \, \xt_0^5} \ln\left( \frac{1 - \xt_0}{1+ \xt_0} \right) \\
& \,
- \frac{1 - 12 \xt_\lambda^2 + 18 \xt_\lambda^4}{6 \, \xt_\lambda^5} \ln\left( \frac{1 - \xt_\lambda}{1+ \xt_\lambda} \right) + \frac{28}{15} \ln\left(\frac{\Lambda^2}{M^2} \right) - \frac{7}{6} \ln ( 32 \pi \breve{G}_\Lambda \Lambda^2 )  - \frac{1}{30} \ln(2) \, . 
\end{split}
\ee
Here we used the subscript on $\xt$ to discriminate $\xt_0 = \xt|_{\beta =0}$ and $\xt_\lambda = \xt|_{\beta = -2 \lambda}$ and substituted $g=\breve{G}_\Lambda \Lambda^2$ to make the dependence on the UV-cutoff explicit. Thus there is no accidental cancellation of the non-local terms when combining the contributions from the scalar, vector, and tensor sector. This also holds when setting $\lambda=0$. This completes our evaluation of the reconstruction formula.
%-----------------------------------------------------------
\bibliographystyle{apsrev4-2}
\bibliography{reconstrbib}
%-----------------------------------------------------------
\end{document}